\documentclass[10pt,twocolumn,twoside]{IEEEtran}
\IEEEoverridecommandlockouts
\usepackage{caption}
\usepackage{subfigure}
\usepackage{cite}
\usepackage{amsmath,amssymb,amsfonts}
\usepackage{algorithmic}
\usepackage{graphicx}
\usepackage{textcomp}
\usepackage{xcolor}
\usepackage{cite}
\usepackage{amsmath,amssymb,amsfonts}
\usepackage{algorithmic}
\usepackage{graphicx}
\usepackage{textcomp}
\usepackage{xcolor}
\usepackage{algorithm}
\usepackage{comment}
\usepackage{setspace}
\usepackage{color}
\usepackage{booktabs}
\usepackage{multirow}
\usepackage{epstopdf}
\usepackage{amsmath}
\usepackage[margin=0.7in]{geometry}

\def\BibTeX{{\rm B\kern-.05em{\sc i\kern-.025em b}\kern-.08em
    T\kern-.1667em\lower.7ex\hbox{E}\kern-.125emX}}
\begin{document}

\title{Spatio-temporal Modeling for Large-scale Vehicular Networks Using Graph Convolutional Networks}
\author{\IEEEauthorblockA{Juntong~Liu\IEEEauthorrefmark{1}, Yong~Xiao\IEEEauthorrefmark{1}\IEEEauthorrefmark{3}, Yingyu Li\IEEEauthorrefmark{1}, Guangming~Shi\IEEEauthorrefmark{2}\IEEEauthorrefmark{3},  Walid Saad\IEEEauthorrefmark{4}, and H. Vincent Poor\IEEEauthorrefmark{5} \\%, Yi Zhong\IEEEauthorrefmark{1}, Tao Han\IEEEauthorrefmark{1}\\
\IEEEauthorblockA{\IEEEauthorrefmark{1}School of Electronic Inform. \& Commun., Huazhong Univ. of Science \& Technology, China}\\
\IEEEauthorblockA{\IEEEauthorrefmark{2}School of Artificial Intelligence, Xidian University, Xi'an, China}\\
\IEEEauthorblockA{\IEEEauthorrefmark{3}Pazhou Lab, Guangzhou, China}\\
\IEEEauthorblockA{\IEEEauthorrefmark{4}Bradley Department of Electrical and Computer Engineering, Virginia Tech, VT}\\
\IEEEauthorblockA{\IEEEauthorrefmark{5}School of Engineering \& Applied Science, Princeton University, Princeton, NJ}
%\thanks{This work was supported in part by the U.S. National Science Foundation under Grant CCF-1908308.}
}
} 
\maketitle

\begin{abstract}
The effective deployment of connected vehicular networks is contingent upon maintaining a desired performance across spatial and temporal domains. In this paper, a graph-based framework, called \emph{SMART}, is proposed to model and keep track of the spatial and temporal statistics of vehicle-to-infrastructure (V2I) communication latency across a large geographical area. \emph{SMART} first formulates the spatio-temporal performance of a vehicular network as a graph in which each vertex corresponds to a subregion consisting of a set of neighboring location points with similar statistical features of V2I latency and each edge represents the spatio-correlation between latency statistics of two connected vertices. Motivated by the observation that the complete temporal and spatial latency performance of a vehicular network can be reconstructed from a limited number of vertices and edge relations, we develop a graph reconstruction-based approach using a graph convolutional network integrated with a deep Q-networks algorithm in order to capture the spatial and temporal statistic of feature map pf latency performance for a large-scale vehicular network. Extensive simulations have been conducted based on a five-month latency measurement study on a commercial LTE network. Our results show that the proposed method can significantly improve both the accuracy and efficiency for modeling and reconstructing the latency performance of large vehicular networks.
\end{abstract}

\vspace{-0.1in}
\section{Introduction}
With the rapidly growing demand on intelligent vehicular services and applications, connected vehicles that rely on external communication, computation, and storage resources to facilitate decision making and driving assistance have become increasingly popular. According to the recent report\cite{ihsmarkit}, in 2025 over 60\% of new vehicles sold globally will be connected to the Internet by wireless technologies such as 5G and beyond\cite{xyee}.
%Over 250 million vehicles on the road are expected to support over-the-air software updating capabilities. 

Despite this surge in popularity, there exists many challenges. In particular, there is a need to better understand how the achievable communication latency over spatial and temporal domains. For instance, due to the heterogeneity in services and applications as well as the diversity of service-requesting devices such as wearable devices\cite{wearable}, sensors\cite{sensor}, LiDar\cite{lidar}, and others, the maximum tolerable latency of different vehicular services can dynamically change across a wide range. Moreover, the latency of a large-scale vehicular network is location-dependent, closely related to the potential signal blockage and interference caused by factors such as the surrounding environment as well as the distribution of the network infrastructure. The challenge for spatial and temporal latency modeling is further exacerbated by the fact that vehicles are consistently moving from one location to another, causing frequent service and link changes. As such, there is a need to develop a simple and effective solution to capture the performance, in terms of latency, of a large-scale vehicular network across different time and location. 

According to recent observation reported in {\cite{adaptivefog}\cite{xiarong}}, the instantaneous latency performance of each mobile device does not exhibit any noticeable spatial and temporal correlations. The statistical feature such as probability distribution function (PDF) however does show strong spatial and temporal dependencies. This makes it natural to develop a graph-based model to capture the statistical features of a vehicular network in which each location point can be seen as a graph vertex and each edge could represent the spatial correlation between two connected location points. Despite its potential, formulating a graphical model to characterize the interactive latency (e.g. round-trip time (RTT)) of a vehicular network faces the following novel challenges. First, it is generally impossible to constantly collect samples across a wide geographical area and keep track of temporal statistics at all locations. Second, the correlation of latency performance at different time stamps and locations can be complex and difficult to measure. There are still lacking commonly adopted metrics to quantify the correlation of the statistical distributions of interactive latency. Finally, vehicles driving at different locations may request different subsets of services, each of which may have unique service demands and requirements. Thus, due to the random nature of wireless networks, it is generally impossible to always support all the requested services with the guaranteed performance.

%插入related works：
\vspace{-0.15in}
\subsection{Related Works}
%\vspace{-0.05in}
Minimizing the latency of communication links is essential for next generation wireless technology\cite{xy6gself}. Most existing works focused on how to maintain the latency experienced by a connected vehicle below a deterministic threshold. In particular, the authors in \cite{adaptivefog} propose AdaptiveFog, a novel framework to maximize confidence levels in LTE-based fog computing for smart vehicles. In \cite{xiarong}, the authors propose a spatio-temporal generative learning  model to reconstruct the missing latency samples based on a variation autoencoder.

There has been a number of recent works \cite{stcommunication,v2vcaching,stgcn} that investigated the spatial and temporal features of a vehicular networks. For examples, in \cite{stcommunication}, the authors investigate the temporal and spatial dynamics of vehicular ad-hoc networks in order to study communication properties such as adaptability, stability, and effectiveness for safety-critical applications, where latency plays a major role. In \cite{v2vcaching}, a novel spatio-temporal caching strategy is proposed based on the notion of temporal graph motifs that can capture spatio-temporal communication patterns in vehicle-to-vehicle networks. In \cite{stgcn}, the authors propose a framework based on spatio-{}temporal graph convolutional networks (GCN) for traffic prediction using spatio-temporal blocks to capture comprehensive spatio-temporal correlations in multi-scale traffic networks.

However, the prior art in \cite{adaptivefog,xiarong,stcommunication,v2vcaching} does not consider the dynamics of the statistical features for large vehicular networks. Therefore, there is a strong need for developing a simple but effective solution that can exploit the spatio-temporal correlation for tracking the latency performance of large-scale vehicular network. 

%\addtolength{\topmargin}{0.28in}
\vspace{-0.18in}
\subsection{Contributions}
%\vspace{-0.05in}
The main contribution of this paper is a novel graphical modeling and reconstruction framework, called {\it SMART} (Spatio-temporal Modeling And ReconsTruction), that can be used to characterize the feasibility of supporting different latency-sensitive services in a vehicular network across a large geographical area. In particular, we first model a large vehicular network as a graph by dividing the entire service area into different subregions, each of which corresponds to a vertex consisting of connected location points with similar latency statistical probabilities. Any two neighboring subregions will be connected with an edge. Statistical distance measures such as the Jensen-Shannon (JS) divergence have been introduced to quantify the correlation between neighboring subregions. {\it SMART} adopts GCN and deep Q-networks (DQN) to capture the latency graphs' spatial and temporal features, respectively. We show that, when some graphical features change, the captured spatial correlation is sufficient to reconstruct the complete updated graphical structure of a large vehicular network from an incomplete set of samples collected from a limited number of subregions. To accelerate the reconstruction speed of a large vehicular network, we propose an efficient graph reconstruction solution based on natural gradient descendant (NGD). We conduct extensive performance evaluation using real traces collected over a five-month measurement campaign in a commercial LTE network. Simulation results show that our proposed method can accurately recover the spatio-temporal latency performance across all the subregions in a large vehicular network.

The rest of the paper is organized as follows. In Section \uppercase\expandafter{\romannumeral2}, we present the preliminary observations. Section \uppercase\expandafter{\romannumeral3} describes the methodology used in our framework in detail. In Section \uppercase\expandafter{\romannumeral4}, we describe experimental setups and present the simulation results. Finally, we conclude the paper and discuss potential future works in Section \uppercase\expandafter{\romannumeral5}.

\vspace{-0.10in}
\section{Preliminary Observation and ARCHITECTURE OVERVIEW}
\vspace{-0.02in}
\subsection{Preliminary Observation}
\vspace{-0.01in}
The latency of wireless communication systems is known to exhibit spatial and temporal variation. Here, we particularly focus on the wireless access latency between a moving vehicle and the first IP address (i.e., the first node encountered in a cellular system) of a commercial LTE network, also called the vehicle-to-infrastructure (V2I) communication latency. 
\vspace{-0.12in}
\begin{figure}[h]
\centering
\includegraphics[width=7cm]{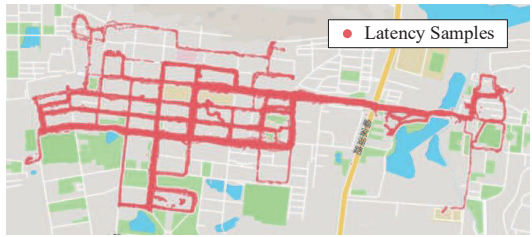}
\caption{\small{Measuring routes and traces of our dataset.}}
\label{fig ditu}
\end{figure}
\vspace{-0.2in}
\begin{figure}[htbp]
\subfigure[]{
  \begin{minipage}[t]{0.45\linewidth}
   \centering
   \includegraphics[width=4.2cm]{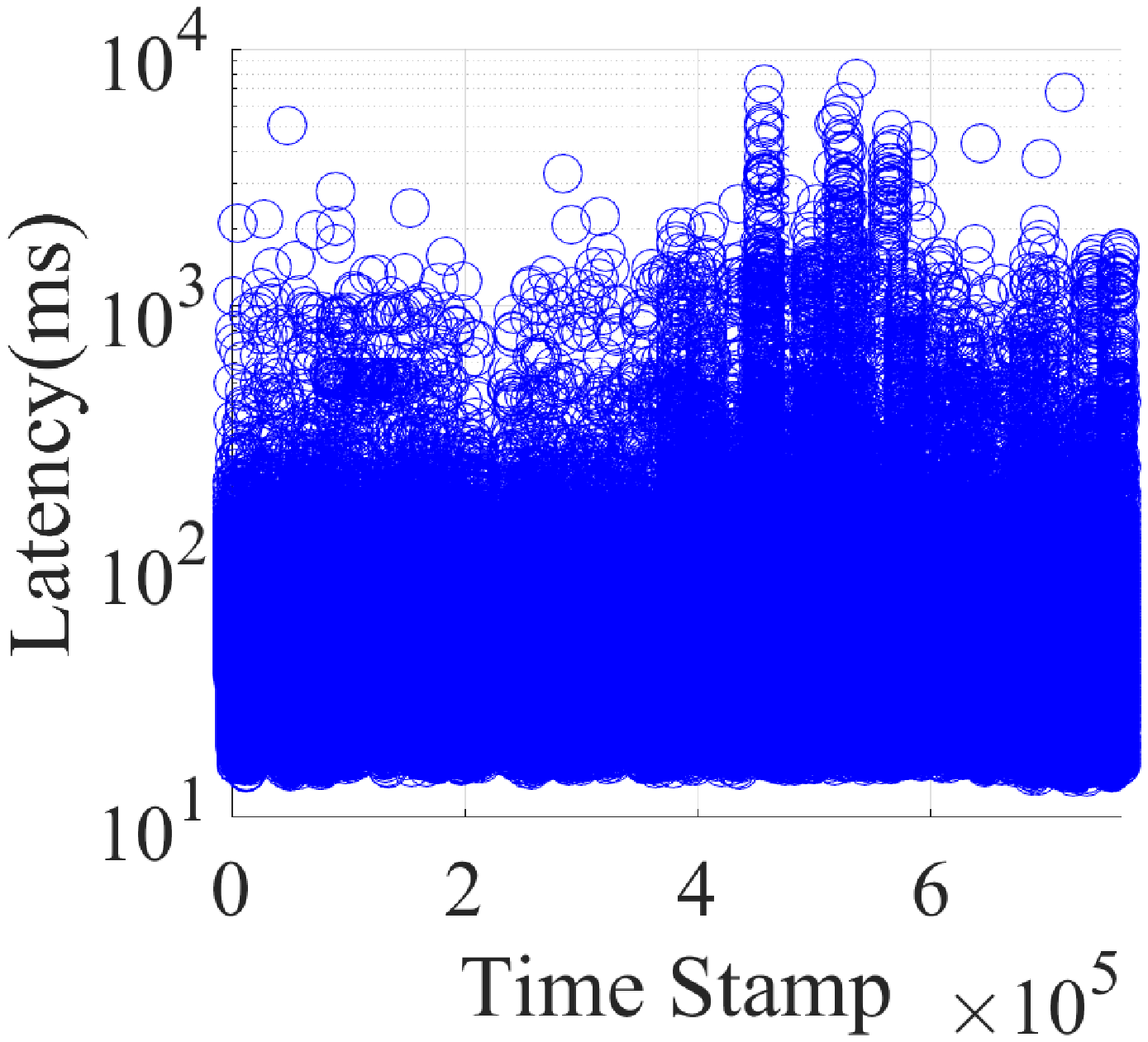}
   %\caption{fig1}
  \end{minipage}%
}
\subfigure[]{
  \begin{minipage}[t]{0.45\linewidth}
   \centering
   \includegraphics[width=4.4cm]{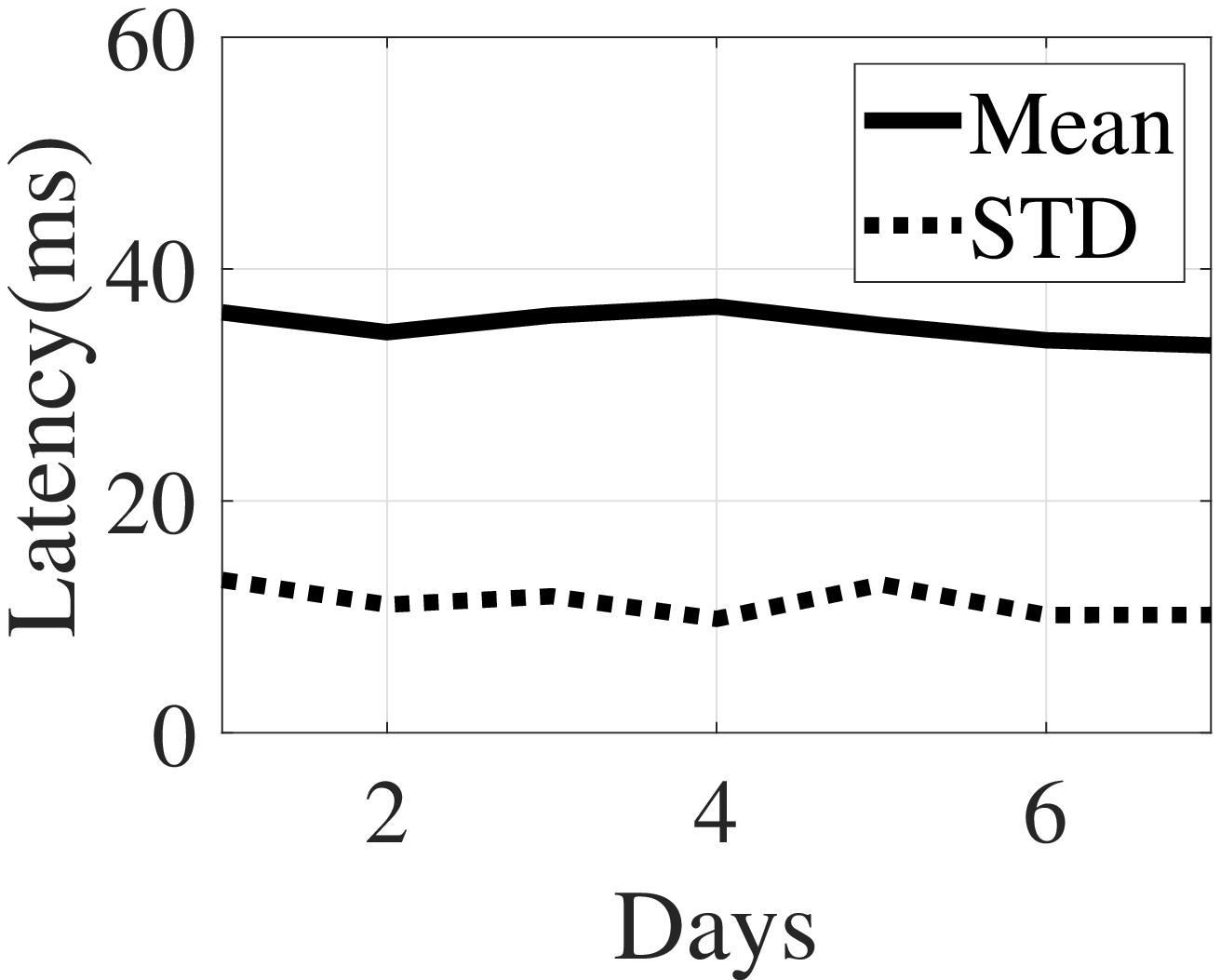}
   %\caption{fig1}
  \end{minipage}%
}
\vspace{-0.15in}
\caption{\small{(a) Latency samples recorded in a university campus listed in the sequence of time stamps. (b) Mean and STD of samples measured through a consecutive week in the same location.}}
\label{fig delay_time}
\end{figure}
\vspace{-0.2in}
\begin{figure}[ht]
%\centering
\subfigure[]{
  \begin{minipage}[t]{0.45\linewidth}
   \centering
   \includegraphics[width=4.2cm]{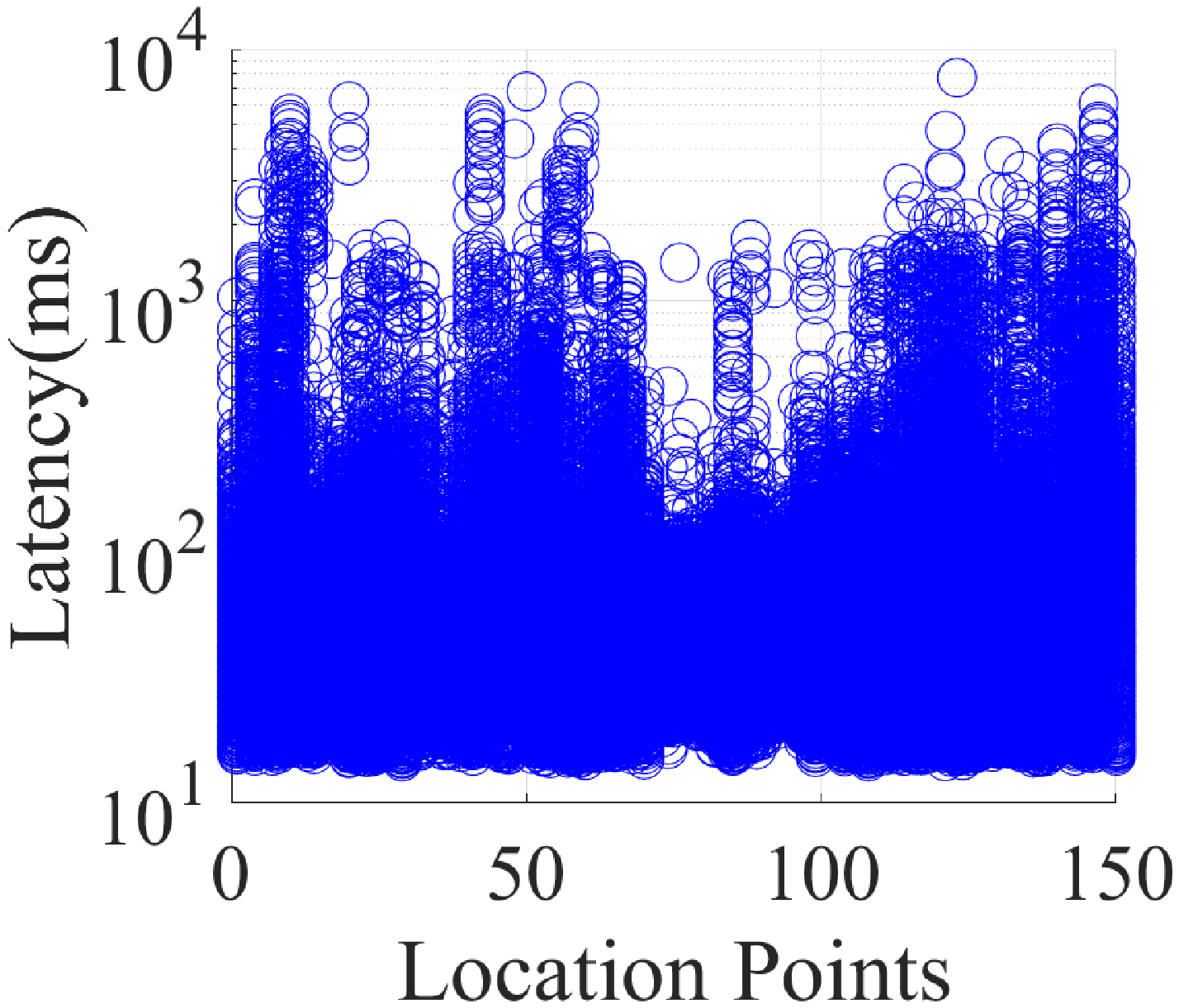}
   %\caption{fig1}
  \end{minipage}%
}
\subfigure[]{
  \begin{minipage}[t]{0.45\linewidth}
   \centering
   \includegraphics[width=4.4cm]{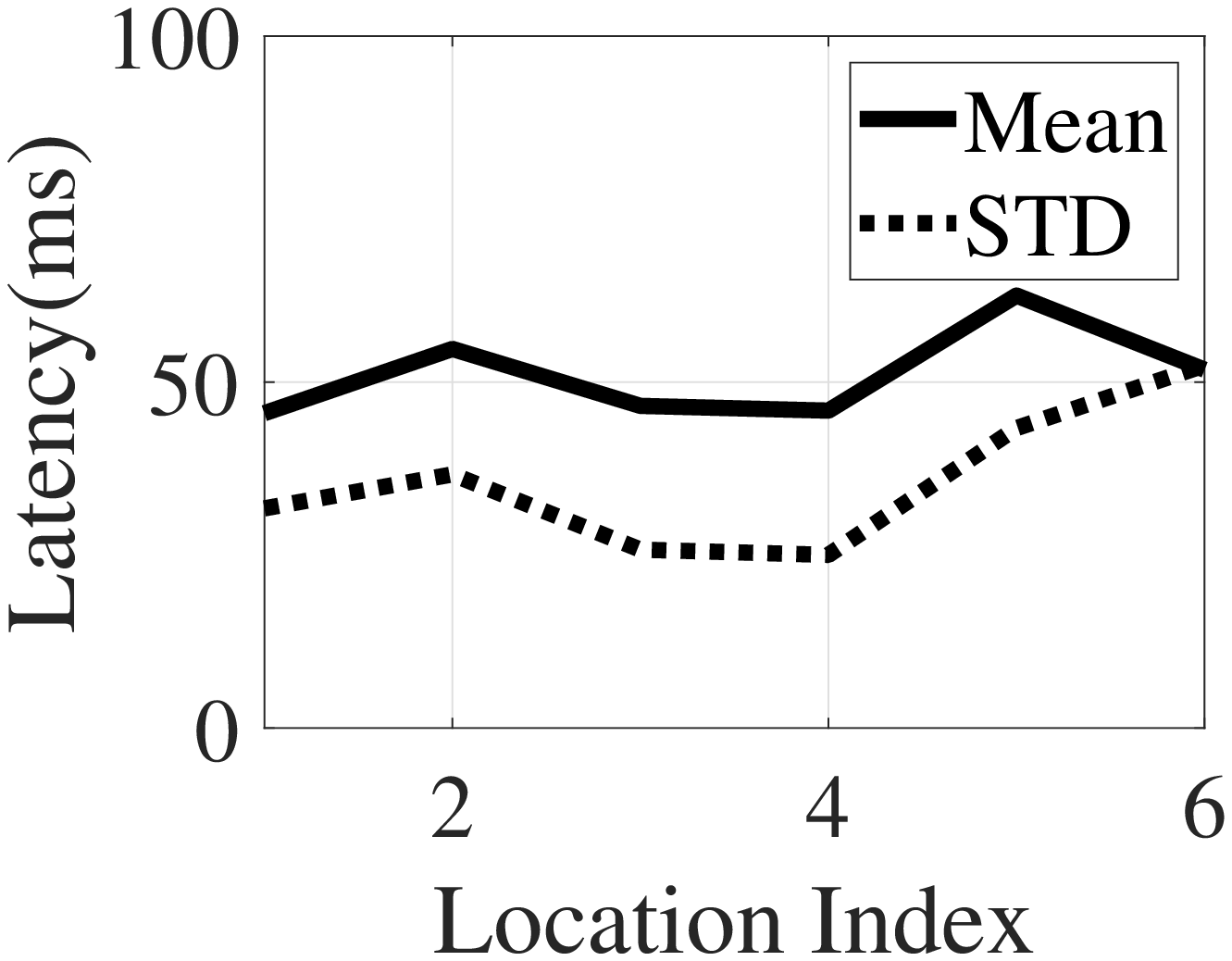}
   %\caption{fig1}
  \end{minipage}%
}
\vspace{-0.15in}
\caption{\small{(a) Latency samples recorded in a university campus listed in the sequence of location points. (b) Mean and STD of samples from 6 locations on a main driving route.}}
\label{fig delay_loc}
\end{figure}
\vspace{-0.15in}

%\vspace{-0.1in}
We adopt a dedicated smart phone app using Android API to periodically ping the first node and record the RTT for both data delivery and acknowledgment. Fig. \ref{fig ditu} shows the measurement routes and traces of our dataset. We consider the RTT as the main metric for interactive wireless access latency for an LTE-supported connected vehicular system. Existing works as well as our own observation have already shown that even two consecutive measurements of the RTT at the same location can vary significantly. In addition, the temporal and spatial correlation of the instantaneous RTTs are often negligible. Fortunately, the statistical features such as mean and standard deviation (STD) remain relatively stationary. In Figs. \ref{fig delay_time} and \ref{fig delay_loc}, we present the mean and STD of RTT samples collected at different location points throughout the main driving route of a university campus (see Fig. \ref{fig delay_loc}(b)) as well as these collected at the same university lab location over a consecutive week (see Fig. \ref{fig delay_time}(b)). We can observe that compared to instantaneous latency samples, the mean and STD vary relatively slow according to different time and location. Also, the spatial variation causes a more noticeable impact on the mean and STD, compared to the temporal variation.

In this work, we plan to exploit the spatial and temporal correlation of the statistics of RTTs and model the vehicular network as a graph in which each vertex corresponds to a specific subregion and the edge connecting two vertices represents the statistical distance of the PDFs of RTTs between two connecting locations.

To characterize the temporal correlation of the graph, we consider a slotted process and assume the graphical model of latency statistics within each time slot can be considered to be fixed. Motivated by the fact that the temporal variation of the statistical features of the RTT at the same location often changes in a much slower pace than the statistical difference between different locations, we adopt a reinforcement learning-based approach to sequentially select a subset of subregions at the beginning of each time slot to collect samples and then exploit a GCN-based approach to reconstruct the update statistical features of all the location points in the new time slot. The reconstructed model will then be evaluated and compared with the real RTTs collected during the rest of the entire time slot. The evaluation results will then be used to update the model in future time slots.

\vspace{-0.15in}
\subsection{Architecture Overview}
\begin{figure}[t]
\centering
\includegraphics[width=8cm]{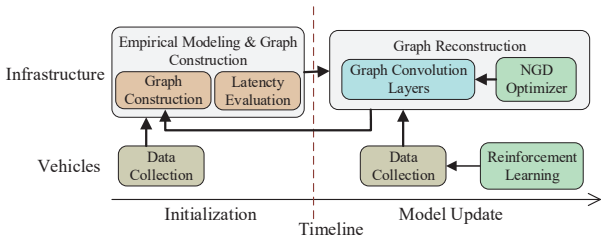}
%\vspace{-0.3in}
\vspace{-0.08in}
\caption{\small{Model Architecture.}}

\label{fig model}
\vspace{-0.2in}
\end{figure}

\vspace{-0.05in}
We propose {\it SMART}, a novel architecture for modeling and keeping track of spatial and temporal statistics of wireless access latency between connected vehicles and wireless infrastructure across a large geographical area. The proposed architecture consists of three major components: data collection, empirical modeling and graphical model construction, model update and reconstruction, as illustrated in Fig. \ref{fig model}. We give a more detailed exposition of each components as follows.

{\bfseries Data Collection:} 
We consider a connected vehicular system in which each vehicle is connected to a commercial LTE network owned by a mobile network operator while driving throughout an area of consideration. The RTTs of data packet delivered to the first wireless infrastructure node as well as receiving the feedback are recorded and reported to a central server. Note that it is not necessary for the central server to constantly collect RTT samples from all the vehicles. The server only needs to request a limited number of RTT samples from vehicles located in a carefully selected subset of regions at the beginning of each time slot.

{\bfseries Empirical Modeling and Graph Construction:} 
As observed in Section \uppercase\expandafter{\romannumeral2}-A, the statistical features of RTTs collected at different locations can exhibit strong spatial correlation. We adopt a statistical distance that can be used to calculate the confidence of the latency performance at each location. We can therefore establish an empirical graphical model.

%\addtolength{\topmargin}{0.2in}

{\bfseries Graph Reconstruction:} 
%Although the latency distribution at one location point presents well periodicity over time, it can vary significantly over one period according to our measurement. 
To deal with the temporal variance and maintain a real-time update of the graphical model, we consider a slotted process. We use a reinforcement learning-based method to select a small subset of locations that collect samples at the beginning of each time slot so as to give more accurate graph reconstruction based on these samples. Then, we exploit a GCN-based approach to reconstruct the confidence of the whole graph vertices in the new time slot from s selected subset of subregions. 

%In Section \uppercase\expandafter{\romannumeral3}
%{\bfseries Performance Evaluation}
\vspace{-0.13in}
\section{Methodology}
\vspace{-0.05in}
%{\bfseries Problem Formulation}:
In this section, we consider a commercial LTE network for connected vehicles across a university campus as an example to describe how to apply {\it SMART} to model and construct a spatial and temporal statistical modeling of latency performance of a V2I network. To make our description clearer, we first introduce the following notations and concepts. We model the roadways in the campus as an undirected graph $G=\langle\mathcal{V}, \mathcal{E}, \boldsymbol{X}\rangle$ where $\mathcal{V}=\left\{v_{1}, v_{2}, \cdots, v_{N}\right\}$ is the set of vertices representing $N$ locations; $e_{ij} \in \mathcal{E}$ is the edge between vertices with the weight characterized by the similarity (JS divergence) of latency distributions between vertex $v_{i}$ and $v_{j}$, which will be discussed more thoroughly later in Section \uppercase\expandafter{\romannumeral3}-B. $\boldsymbol{X}=\left [\mathbf{x}_{1}, \mathbf{x}_{2}, \cdots, \mathbf{x}_{N}\right] \in \mathbb{R}^{N \times F}$ is the vertex feature matrix of the graph and $F$ is the dimension of the feature vector of each vertex. $\boldsymbol{A} \in \mathbb{R}^{N \times N}$ is the adjacent binary matrix of graph $G$, i.e., for any $v_{i}, v_{j} \in V$, $A_{i j}=A_{j i} \in\{0,1\}$ where $1$ means that there is an edge between $v_{i}$ and $v_{j}$ and $0$ otherwise. $\boldsymbol{D}$ is the degree matrix whose element is $D_{i i}=\sum_{j} A_{i j}$.
%, which is calculated through JS divergence; 

\vspace{-0.15in}
\subsection{Data Collection}
\vspace{-0.02in}
Data collection will be conducted at the beginning of both initialization and the model updating process as shown in Fig. \ref{fig model}. We adopt a smartphone app, called {\it Delay Explorer}, that periodically pings the first node and record RTTs every $500$~ms. It can also record data such as time stamps, GPS coordinates, and driving speed, among others. We next explain, in detail, the data collection procedure of data collection and model updating processes: 

{\it 1)} During the initialization process, RTT samples will be first collected throughout each considered area. The latency data is collected constantly until there are enough samples for each location to establish an empirical PDF. The PDFs will then be used in initial graph construction and latency performance evaluation.

{\it 2)} During the model updating process, we consider a slotted process to cope with the temporal variation of latency performance and update the model slot by slot. At the beginning of each time slot, the model will carefully choose a subset of locations to collect instantaneous RTT samples as will be discussed in Section \uppercase\expandafter{\romannumeral3}-C. 
We will evaluate the latency performance of the selected locations as well as their labels based on these data in the graph reconstruction procedure. %They can then be the training dataset in the semi-supervised GCN-based method.
%These data will be calculated into confidence of the performance in the graph reconstruction component and offer service level prediction as the labeled data in a semi-supervised GCN-based method. The length of time slot is set to be 24 hours in our simulation in which the data will be collected constantly. At the end of each time slot, the latency samples will be collected adequately to acquire the real latency performance in every location points.

\vspace{-0.15in}
\subsection{Empirical Modeling and Graph Construction}
\vspace{-0.02in}
After collecting a sufficient number of samples, we can establish an empirical PDF for each location vertex in the graph. We adopt JS divergence to characterize the statistical correlation between two neighboring locations, i.e., weight of the edge connecting two vertices. Let $P_{i}(x)$ and $P_{j}(x)$ be the PDFs of latency at location $i$ and $j$, respectively. The JS divergence can be written as follows:
\begin{eqnarray}
\textrm{JS}\left(P_{i}(x) \| P_{j}(x)\right)=\frac{1}{2} \textrm{KL}\left(P_{i}(x) \| \frac{P_{i}(x)+P_{j}(x)}{2}\right)\nonumber\\ +\frac{1}{2} \textrm{KL}\left(P_{j}(x) \| \frac{P_{i}(x)+P_{j}(x)}{2}\right),
%\textrm{KL}(P_{i}(x) \| P_{j}(x))=\sum_{x \in X} P_{i}(x)\log \frac{P_{i}(x)}{P_{i}(x)},
\end{eqnarray}
where $\textrm{KL}(P_{i}(x) \| P_{j}(x))=\sum_{x \in X} P_{i}(x)\log \frac{P_{i}(x)}{P_{i}(x)}$. If the JS divergence $\textrm{JS}\left(P_{i}(x) \| P_{j}(x)\right)$ falls belows a pre-defined threshold $\eta$, we can add an undirected link between locations $i$ and $j$.
%这里要补充一下权值的设置
In this way, we can then model the campus roadway network as an undirected graph.

As already discussed, we consider the time-varying statistical features of RTTs within a slotted time duration. The statistical feature we mainly focus on is the confidence, defined as the probability of a certain latency threshold can be satisfied, of the latency performance at each location in a multi-service scenario. We divide the locations into different classes according to their confidence for supporting some key vehicular services and give each location a preliminary label. To make our model more general, the number of classes could be flexibly assigned so that the evaluation can be adopted to various scenarios. We use the following statistical requirement as the latency performance metric: 
\vspace{-0.08in}
\begin{eqnarray}
{\rm Pr}(d_{i}\leq \tau)=\int_{0}^{\tau}{P_i\left(x\right)d}x \leq \varepsilon,
\label{eq confidence}
\end{eqnarray}
\vspace{-0.03in}
where ${\rm Pr}(\cdot)$ represents the probability of an event, and $\varepsilon$ is the maximum confidence that must be guaranteed at the $i$th location. $\tau$ represents maximum tolerable latency for some specific vehicular services. According to the 5GAA\cite{5gaa}, the latency requirement of major vehicular services can be roughly classified into 5 major use cases listed in Table \ref{tab usecase}. The latency requirements of these services can be roughly divided into three classes according to Eq. (\ref{eq confidence}):
\begin{enumerate}
\item Service Level 1: $\tau$ is $100$~ms and $\varepsilon$ is $99.99\%$;
\item Service Level 2: $\tau$ is $100$~ms and $\varepsilon$ is $99\%$;
\item Service Level 3: $\tau$ is $120$~ms and $\varepsilon$ is $99\%$.
\end{enumerate}

It should be noticed that the above three service levels exhibit an inclusion relation: service satisfying level 1 (or level 2) requirement can also meet the requirement of level 2 (or level 3). For example, a location in Level 2 could satisfy the service requirement of a hazardous location warning, but it cannot satisfy the needs of a intersection assistance movement at a crossroad.
%\vspace{-0.07in}
\begin{table} [t]
\centering
\footnotesize
\caption{Latency Requirement of 5 Different Services} 
\setlength{\tabcolsep}{2mm}{
\renewcommand\arraystretch{1.5}
\begin{tabular} {|c|c|c|} 
\hline{}
\textbf{Service Type} & \textbf{Service Level Latency} & \textbf{Reliability}\\ 
\hline 
Intersection movement & $100$~ms & 99.99\% \\
\hline
Awareness of the presence & \multirow{2}*{$100$~ms}&\multirow{2}*{99.9\%}\\
of vulnerable road user & ~ & ~\\ 
\hline
Hazardous location warning & $100$~ms & 99\%\\
\hline
Cross-traffic left-turn assist & $100$~ms & 90\% \\
\hline
Emergency break warning & $120$~ms & 99\% \\
\hline
\end{tabular}}
\label{tab usecase}
\vspace{-0.25in}
\end{table}

%confidence of latency divide into three levels [5gaa] 
\vspace{-0.13in}
\subsection{Graph Reconstruction}
In a practical system, latency performance can be time-varying as shown in Section \uppercase\expandafter{\romannumeral2}. Thus, we need to keep the entire graph updated whenever some locations' latency performance as well as their supported service level change. To characterize the temporal correlation of the graph, we consider a slotted process and assume that the graphical model within each time slot can be assumed to be fixed. Due to the location proximity, the latency variation of two neighboring locations may experience similar changing patterns. We then introduce a GCN-based approach to recover the complete graph from a limited number of RTT samples collected throughout a subset of edges and vertices%The detailed progress is described as follows:

\subsubsection{\it Graph Convolutional Network}

The standard convolution in CNN is not applicable to graphs due to their non-Euclidean structure. We therefore adopt GCN to reconstruct the graphical model whenever some parts of the graph change. According to \cite{semigcn}, the computational complexity of spectral GCN is $O\left(n\right)$ where $n$ is the number of graph edges. Hence, GCN-based approach can be directly applied into large graphs.

%We first introduce the GCN which is used to in the model updating process.  We use the approach in \cite{semigcn} to apply convolutions to graph. Previous works have reduced the computational complexity of spectral GCN from $\mathcal{O}\left(n^2\right)$  to a linear complexity in the number of graph edges\cite{chebyshev}\cite{semigcn}. Hence this approach can be applied into large graphs.

The propagation process of the stacking layers can be written as:

%The normalized Laplacian matrix is $L=I_{N}-D^{-\frac{1}{2}}AD^{-\frac{1}{2}}=U \Lambda U^{T} \in \mathbb{R}^{N \times N}$. $U \in \mathbb{R}^{N \times N}$ is the matrix of eigenvectors of $L$ and $\Lambda \in \mathbb{R}^{N \times N}$ is the diagonal matrix of eigenvalues of $L$.

%By approximating localized graph convolutional layers with the first-order Chebyshev polynomial\cite{semigcn}, the propagation process of the stacking layers can be write as:
\vspace{-0.2in}
\begin{eqnarray}
H^{(0)}=\boldsymbol{X}\  \mbox{and}\  H^{(l+1)}=\sigma(\Delta \boldsymbol{A}H^{(l)}W_{l}),
\label{eq layer}
\end{eqnarray}

where $\Delta \boldsymbol{A}=\tilde{\boldsymbol{D}}^{-\frac{1}{2}} \tilde{\boldsymbol{A}} \tilde{\boldsymbol{D}}^{-\frac{1}{2}}$ is a renormalized matrix with $\tilde{\boldsymbol{A}}=\boldsymbol{A}+\boldsymbol{I_{N}}$ and $\tilde{D}_{i i}=\sum_{j} \tilde{A}_{i j}$. $H^{(l)}$ is the output of layer $l$, $\sigma(\cdot)$ is an activation function which is normally set as ReLU function. $W_{l}$ is the learnable parameter matrix which can be obtained using gradient descent. The input feature vector $\mathbf{x}_i$ can be some normalized latency samples of location point $i$.

To accomplish the semi-supervised classification task, we use the softmax activation function, defined as ${\rm softmax}(\mathbf{x}_i)={\rm exp}(\mathbf{x}_i)/\sum_i{\rm exp}(\mathbf{x}_i)$, on the output $Z$ of the last convolutional layer and the cross-entropy error and the loss function $L$ can be formulated as:
\vspace{-0.03in}
\begin{eqnarray}
Z^{'}={\rm softmax}(Z),\ \  \\
L=-\sum_{l\in y_l}\sum_{f=1}^{F}{Y_{{\it lf}}{\rm ln}}{Z^{'}}_{{\it lf}}.
\label{eq layer}
\end{eqnarray}
\vspace{-0.07in}

${Z^{'}}_{{\it lf}}$ is entry $f$ of the vertex's hidden representation labeled $l$. $Y_{{\it lf}}$ is the ground truth of the corresponding label.

%Neural networks are often trained using stochastic gradient descent (SGD)-based methods.  
In order to accelerate the convergence speed of GCN for a large graph, we introduce a second-order gradient descent method called NGD\cite{ngd} to optimize the parameter matrices in the training process of GCN. NGD transforms gradients into so-called {\it natural gradients} that have proved to be much faster compared to the stochastic gradient descent (SGD). %Natural gradient is defined as follows:
%\begin{eqnarray}
%r(\boldsymbol{\theta}) \approx r\left(\boldsymbol{\theta}_{0}\right)+\mathbf{g}_{0}^{\top}\left(\boldsymbol{\theta}-\boldsymbol{\theta}_{0}\right)+\frac{1}{2}\left(\boldsymbol{\theta}-\boldsymbol{\theta}_{0}\right)^{\top} H_{0}\left(\boldsymbol{\theta}-\boldsymbol{\theta}_{0}\right),
%\label{eq ng}
%\end{eqnarray}
%where $H_{0}=H\left(\boldsymbol{\theta}_{0}\right)$ denotes the Hessian matrix at \boldsymbol{\theta}_{0}
%It has been proved in \cite{ngdingcn}, the Hessian matrix of NGD can be approximated by Fisher information, i.e., we have  
Recently, the work in \cite{ngdingcn} used NGD for a semi-supervised classification task in GCN, and it showed encouraging results in both accuracy and convergence speed on some benchmark datasets.

Preconditioning is inspired by the idea that capturing the relation between the gradient of parameters before optimization will help with convergence. For example, the traditional optimizer, such as Adam\cite{adam}, uses diagonal preconditioner which neglects the pair-wise relation between gradients. However, any extra information about gradients is often impossible or hard to obtain. Motivated by NGD, we introduce a preconditioning algorithm that uses the second moment of gradient to approximate the parameters' Fisher information matrix in the prediction distribution\cite{ngdingcn}.

Algorithm \ref{Algorithm 1} shows the detailed preconditioning process for modifying gradients of each layer at any iteration. The gradients are first transformed using two matrices, $\boldsymbol{V}_{l}^{-1}$ and $\boldsymbol{U}_{l}^{-1}$, then sent to the optimization algorithm for parameter updating. Let $m$ be the number of the network layers and $\odot$ be element-wise multiplication operation. 
%$\phi_{l}^{\prime}(\cdot)$ is an nonlinear function like RELU
$\mathrm{x}_{l-1, i}$ represents the output feature vector of $v_i$ in layer $l-1$ and is updated into $\tilde{\mathrm{x}}_{l-1, i}$ using a renormalization trick for $i=\{1,\cdots ,N\}$. $\lambda$ is a hyper-parameter that controls the cost of predicted labels and $\epsilon$ is a regularization hyper-parameter to evaluate $\boldsymbol{V}_{l}^{-1}$ and $\boldsymbol{U}_{l}^{-1}$. 
%\vspace{-0.07in}

\begin{algorithm}[t]
\caption{Preconditioning using NGD}\label{Algorithm 1}
\hspace*{0.02in} {\bf Input:}
Gradient of parameters $\nabla W_{l}$ for $l=1,...,m$, adjacency matrix $\boldsymbol{A}$, degree matrix $\boldsymbol{D}$, training mask $z$, regularization hyper-parameters $\lambda$,$\epsilon$
\begin{algorithmic}[1]
%\STATE $n=\operatorname{dim}(\mathrm{z})$, $\bar{n}=\sum(\mathrm{z})$, $\Delta \boldsymbol{A}==\left[\Delta {a}_{i j}\right]$
\STATE Derive the numbers of labeled and unlabeled vertices via $\bar{n}=\sum(\mathrm{z})$ and $n=\operatorname{dim}(\mathrm{z})$. And let $\left[\Delta {a}_{i j}\right]$ represent the entry of $\Delta \boldsymbol{A}$.

\FOR{$l=1,\cdots,m$}
%\STATE $\tilde{\mathrm{x}}_{l-1, i}=\sum_{j=1}^{n} \Delta{a}_{i, j} \mathrm{x}_{l-1, j}$
\STATE Formulate the feature aggregation process of each layer via $\tilde{\mathrm{x}}_{l-1, i}=\sum_{j=1}^{n} \Delta{a}_{i, j} \mathrm{x}_{l-1, j}$.

%\STATE $\mathbf{u}_{l-1, i}=\partial L / \partial \mathbf{x}_{l} \odot \sigma_{l}\left(W_{l} \tilde{\mathbf{x}}_{l-1, i}\right)$
\STATE  Approximate matrices $\boldsymbol{V}_{l}$ and $\boldsymbol{U}_{l}$ via: \\
$\mathbf{u}_{l-1, i}=\partial L / \partial \mathbf{x}_{l} \odot \sigma_{l}\left(W_{l} \tilde{\mathbf{x}}_{l-1, i}\right)$, \\
$\boldsymbol{U}_{l}=\sum_{i=1}^{n}\left(z_{i}+\left(1-z_{i}\right) \lambda\right) \mathbf{u}_{l-1, i} \mathbf{u}_{l-1, i}^{\top} /(n+\lambda \bar{n})$, \\
$\boldsymbol{V}_{l}=\sum_{i=1}^{n}\left(z_{i}+\left(1-z_{i}\right) \lambda\right) \tilde{\mathbf{x}}_{l-1, i} \tilde{\mathbf{x}}_{l-1, i}^{\top} /(n+\lambda \bar{n})$.
%\STATE $\boldsymbol{U}_{l}=\sum_{i=1}^{n}\left(z_{i}+\left(1-z_{i}\right) \lambda\right) \mathbf{u}_{l-1, i} \mathbf{u}_{l-1, i}^{\top} /(n+\lambda \bar{n})$
%\STATE $\boldsymbol{V}_{l}=\sum_{i=1}^{n}\left(z_{i}+\left(1-z_{i}\right) \lambda\right) \tilde{\mathbf{x}}_{l-1, i} \tilde{\mathbf{x}}_{l-1, i}^{\top} /(n+\lambda \bar{n})$
\STATE {\bf Output:} $(\boldsymbol{V}_{l} + \epsilon^{-1/2}\boldsymbol{I})^{-1} \nabla W_{l} (\boldsymbol{U}_{l} + \epsilon^{-1/2}\boldsymbol{I})^{-1}$
\ENDFOR%\\ \ \\
%\STATE {\bf function} INVERSE(X)
%\STATE \ \ \ {\bf Output:}$(X + \epsilon^{-1/2}I)^{-
\end{algorithmic}
\end{algorithm}

\subsubsection{\it Deep Q-Networks}
As mentioned earlier, the latency performance (i.e. the label of vertices in graph) of each location point can change at different time slots. Always collecting sufficient numbers of samples across all the possible locations is generally impossible. Due to the spatial correlation of the latency graph, collecting a subset of locations will be sufficient to reconstruct the complete updated latency graph. Motivated by the fact that the reconstruction accuracy of a graph model can vary significantly with different sets of selected vertices, in the rest of this section, we formulate the vertices selection for graph reconstruction as a Markov decision process (MDP) defined as follows.

%To keep updating the graphical model when a new time slot arrives, we randomly choose around 10\% nodes in the graph and send the requests to vehicles within them and collect their latency samples. These data are then sent to the semi-supervised learning method mentioned above to reconstruct the graph.
%an intuitive idea in the semi-supervised learning method mentioned above is to randomly choose around 10\% of the nodes in the graph and send the requests to vehicles within them for collecting their latency samples with the first node. 
%However, the prediction accuracy of the trained model can vary significantly with different sets of selected nodes. 
%We consider the timely change of node labels as a hidden markov model (HMM). According to the historical data we can obtain the transition probability $P(s^{\prime}|s)$ from state $s$  $s^{\prime}$. 

%To maximize the prediction accuracy by sequential node selection, we introduce a deep Q-learning (DQN) method\cite{dqn} which is usually chosen to find the optimal policy for sequential decision problems. We first formulate the problem as a Markov decision process (MDP) with finite horizon consisting of the following components:

\noindent {\bfseries State Space $S$} is a finite set of possible service levels that can be supported at each location. $\boldsymbol{S} _t \in \mathbb{R}^{K \times N}$ is a $K \times N$ matrix in time slot $t$ where column vector $\boldsymbol{S}_{t}^{i}$ represents the probability for each $K$ labels in next time slot $t+1$ of vertex $v_i$. Both prior and conditional probability can be obtained from historical data.\footnote{We use a one-month latency collection dataset measured at a university campus. For example, we divide the latency data $d_i$ at location $i$ into $\{d_i^1, d_i^2, \cdots, d_i^p\}$ sequentially according to $p$ time slots and set the label in each slot based on the confidence of latency performance. We can then obtain the conditional probability $\textrm{Pr}(c_j^{(t+1)}|c_i^t)$ for each location where $c_i^t$ means the location is in label $c_i$ at time slot $t$.}

%After the nodes selection, we can derive a graph embedding vector by simply aggregating the node embeddings through trained model. $s_t$= is the state in time slot $t$

\noindent {\bfseries Action Space $A_v$} is the possible selection of location subsets for requesting latency samples. We write $a_v^t=\{v_{1},v_{2},\cdots , v_{m}\}$ as an instance of action in time slot $t$ for $a_v \in A_v$. $\{v_{1},v_{2},\cdots , v_{m}\}$ are $m$ vertices selected from all $N$ vertices in the graph.

\noindent {\bfseries State Transition function $T: S \times A_v \times S \rightarrow[0,1]$} denotes the probability of state transiting from one state to another. When the selection of $m$ vertices in slot $t$ (i.e., the action $a_v^t$ ) is determined, we can observe the actual label $c_i^t$ for each $m$ locations. We can then establish as a mapping function $f:\boldsymbol{S}_{t+1}=f(\boldsymbol{S}_t,a_v^t)$, where the $m$ column vectors with probability $[\textrm{Pr}(c_1^{(t+1)}|c_i^t),\textrm{Pr}(c_2^{(t+1)}|c_i^t),\cdots,\textrm{Pr}(c_K^{(t+1)}|c_i^t)]^{T}$ for state $S_t$ to transit into $S_{t+1}$.
%Then we can replace the $m$ column vectors with $[\textrm{Pr}(c_1^{(t+1)}|c_i^t),\textrm{Pr}(c_2^{(t+1)}|c_i^t),\cdots,\textrm{Pr}(c_K^{(t+1)}|c_i^t)]^{T}$ so that state $S_t$ will transit into $S_{t+1}$. This process can be denoted as a mapping function  $f:\boldsymbol{S}_{t+1}=f(\boldsymbol{S}_t,a_v^t)$.

\noindent {\bfseries Reward Function $R$}: We try to maximize the graph construction accuracy, defined as the percentage of the correctly predicted vertices among all the reconstructed graph, i.e., we have $R_t(\boldsymbol{S}_t,a_v^t)=\frac{1}{N-m}\sum_{v\in \mathcal{V}\backslash a_v^t}{\cal I}\left(c(v)=c_v\right)$ where $c(v)$ is the predicted label and $c_v$ is the true label of vertex $v$. $\cal{I}(\cdot)$ is an indicator function to count the correct prediction.
%label prediction accuracy of unselected locations. So we can write the reward function of prediction accuracy over unselected $N-m$ nodes as $R_t(\boldsymbol{S}_t,a_v^t)$.

We focus on maximizing the long-term reconstruction accuracy including both current and future rewards defined as $Q(\boldsymbol{S}_t,a_v^t)$ when action $a_v^t$ is taken at state $\boldsymbol{S}_t$:
%To maximize the long-term reward, we consider both current and future rewards and define the value function $Q(\boldsymbol{S}_t,a_v^t)$ when action $a_v^t$ is taken at state $\boldsymbol{S}_t$ as:
\begin{eqnarray}
Q(\boldsymbol{S}_t,a_v^t)=R_t(\boldsymbol{S}_t,a_v^t)+\beta Q(\boldsymbol{S}_{t+1},a_v^t),
\label{eq qvalue}
\end{eqnarray}
where $\beta$ is the learning rate.

Following the standard procedure of DQN, we can write the optimal policy $\pi^*$ as:  
%Hence, the optimal policy $\pi^*$ can be write as:
\begin{eqnarray}
\pi^*=\mathop{\arg\min}_{a_v^t\in A_v^{\boldsymbol{S}_t}} Q(\boldsymbol{S}_t,a_v^t)
\label{eq optimalpolicy}
\end{eqnarray}
where $Q(\boldsymbol{S}_t,a_v^t)$ can be pre-calculated and pre-stored in a look-up table (i.e., Q-table) for finding the expected reward under all possible state and action pairs which needs to be stored in each time slot which leads to enormous storage and computational complexity. To address the above problems, DQN uses deep neural networks to estimate the Q-table. The transition $(\boldsymbol{S}_t,a_v^t,R_t(\boldsymbol{S}_t,a_v^t),\boldsymbol{S}_{t+1})$ is stored in the experience relay pool for learning process. During each epoch, the predict network will choose an action which will be evaluated in the target network.
%and it consists of four main parts: \emph{feature input}, \emph{experience replay pool}, \emph{predict network} and \emph{target network}.
%Value of each possible action for each state will be evaluated in predict network and the best action will be chosen with certain probability. Target network will evaluate the value of next state after state transition. 

\vspace{-0.16in}
\section{Simulation Results and Analysis}
\vspace{-0.06in}
In this section, we evaluate the performance of {\it SMART} through extensive simulations using the dataset collected in a university campus. Our simulations are performed mainly using two open-source Python libraries, Pytorch and Pytorch Geometric, on a workstation with an Intel(R) Core(TM) i9-9900K CPU@3.60GHz, 64.0 GB RAM@2133 MHz, 2 TB HD, and two NVIDIA Corporation GP102 [TITAN X] GPUs. 

We consider 150 subregions across the university campus and randomly choose 30 samples in each subregion to construct the feature vector for each vertex. We train GCN models in 200 epochs (training iterations) using both Adam\cite{adam} and NGD with learning rate 0.01. The Adam is used with the weight decay of $5 \times 10^{-4}$ and the momentum of 0.9. A 2-layer GCN with a 16-dimension hidden variable is used in all simulations. The first layer is followed by a drop out function at the rate of 0.5. The training process stops if the validation loss (i.e., the value of loss function on validation set which is used to determine the hyper-parameters in the model) does not decrease for 10 consecutive epochs and the loss function is evaluated using the negative log-likelihood in equation. (5). 

\begin{figure}[t]
%\centering
\subfigure[]{
  \begin{minipage}[t]{0.4\linewidth}
   \centering
   \includegraphics[width=3.7cm]{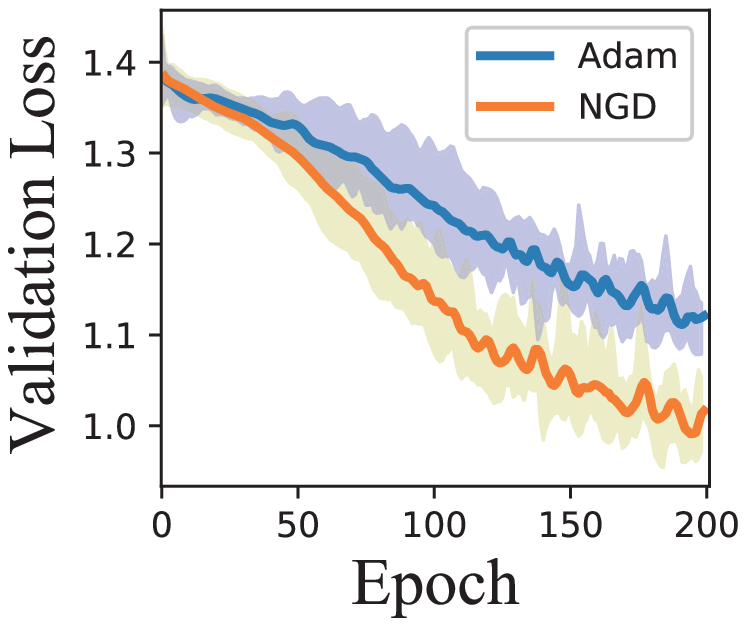}{}
   %\caption{fig1}
  \end{minipage}%
}
\subfigure[]{
  \begin{minipage}[t]{0.5\linewidth}
   \centering
   \includegraphics[width=4.4cm]{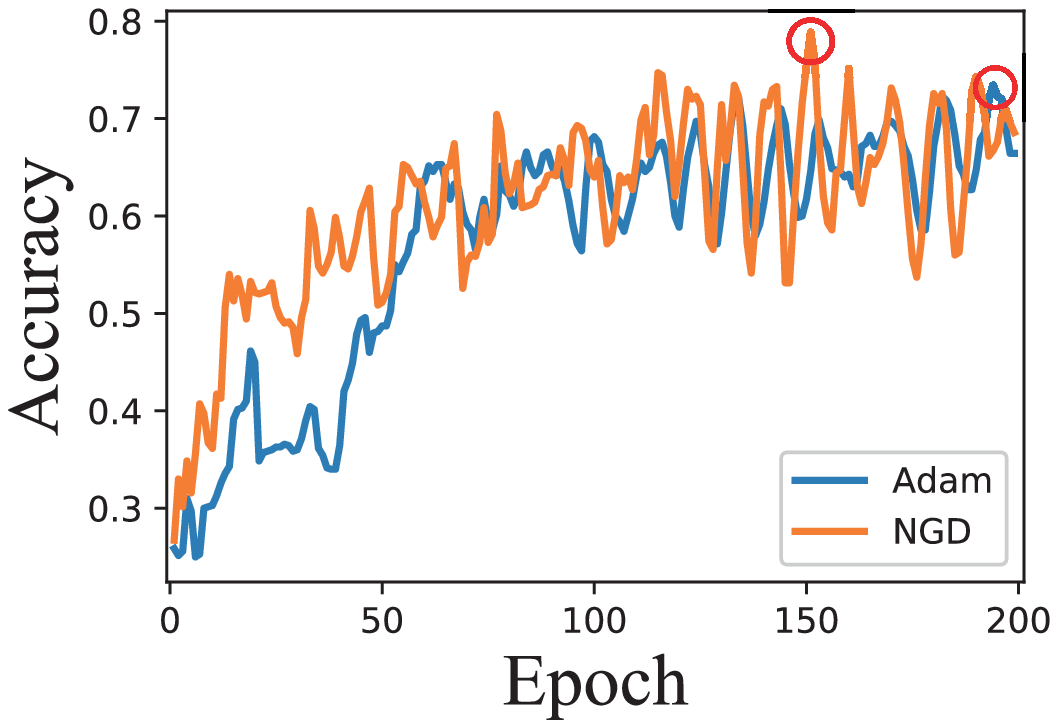}
   %\vspace{-0.14in}
   %\caption{fig1}
  \end{minipage}%
} \vspace{-0.12in}\\
\subfigure[]{
  \begin{minipage}[t]{0.4\linewidth}
   \centering
   \includegraphics[width=3.5cm]{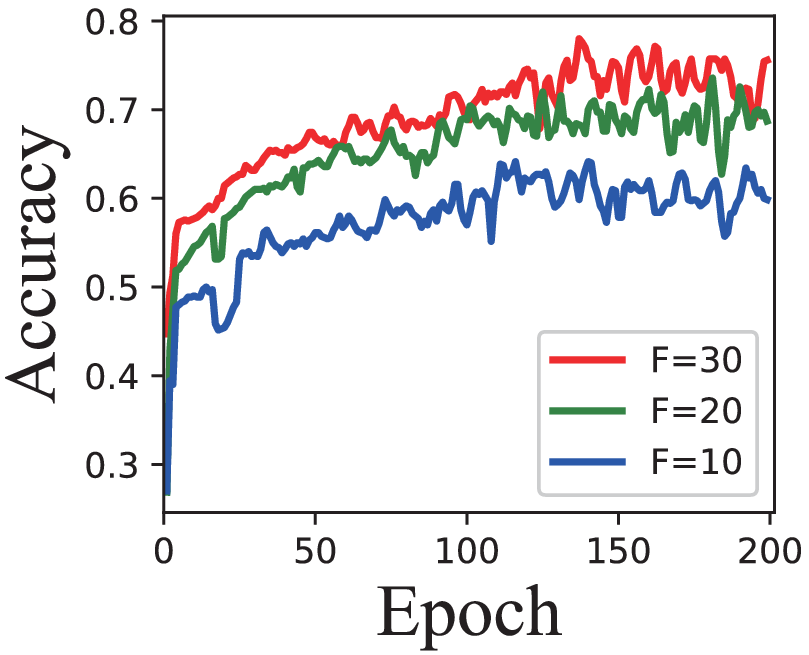}
   %\vspace{-0.14in}
   %\caption{fig1}
  \end{minipage}%
}
\subfigure[]{
  \begin{minipage}[t]{0.5\linewidth}
   \centering
   \includegraphics[width=4.8cm]{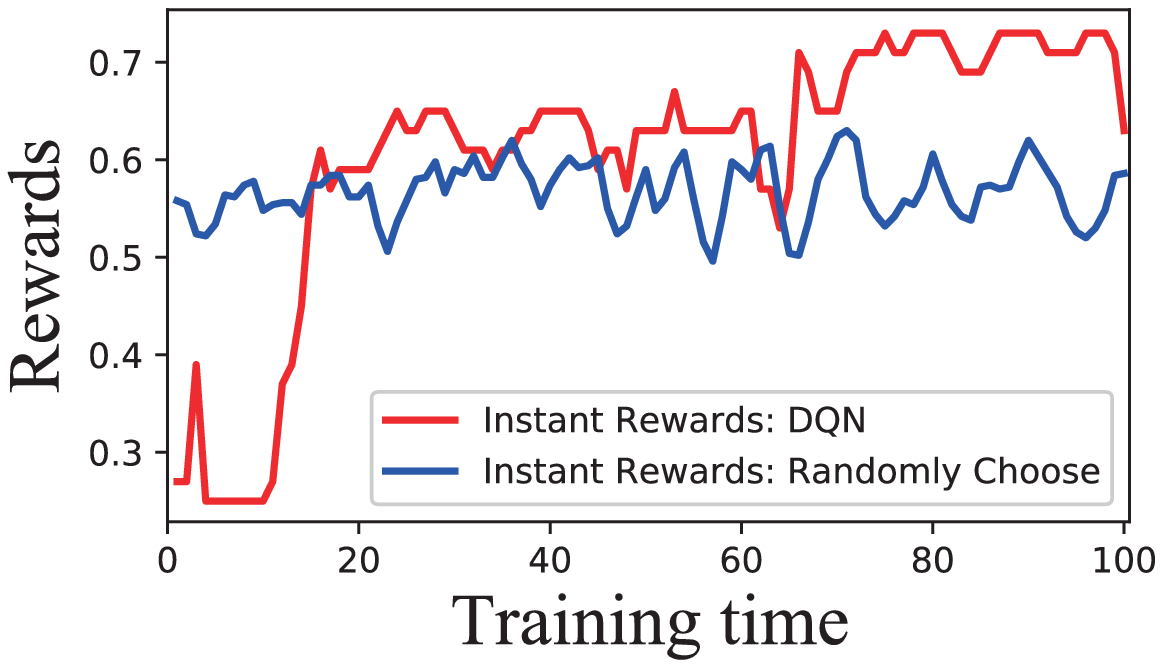}
   %\vspace{-0.14in}
   %\caption{fig1}
  \end{minipage}%
}
\vspace{-0.12in}
\caption{\small{(a) Validation costs. (b) model accuracy based on test dataset, (c) model accuracy with different input feature vectors, and (d) rewards under different number of training iterations.}}
\vspace{-0.22in}
\label{fig result}
\end{figure}

We compare the validation loss and testing accuracy of two optimization methods over 10 rounds in Fig. \ref{fig result}(a). The blue and yellow zones are confidence intervals of Adam and NGD, respectively. We can observe that the validation loss in NGD drops faster than Adam and could result in a lower validation loss. In Fig. \ref{fig result}(b), we compare the model accuracies of Adam and NGD based on our testing dataset. We can observe that GCN with an NGD optimizer can reach a maximum accuracy at 78.86\% over 200 epochs (the red circles in the figure) which outperforms the maximum accuracy of Adam optimizer at 74.43\%. This result demonstrates that the NGD offers faster processing speed compared to traditional optimizer such as standard SGD to reconstruct graph model.

We present model accuracy with different numbers of dimensions $F$ of the input feature vectors. We can observe that the larger dimensions of feature vectors could results in a higher accuracy as shown in Fig. \ref{fig result}(c). 
%It should be noticed that as the input samples increasing, the accuracy grows rapidly at first and then slows down. 
This is because the feature vector with respect to latency samples can offer more information about the latency performance. Fig. \ref{fig result}(d) presents the rewards achieved by the actions selected by DQN under different number of training iterations. We can observe that the reward achieved by selecting locations decided by DQN is always higher than that achieved by adopting random selection of locations at each time slot.

%(fig 3b accuracy of using reinforcement learning vs randomly choose vs choose the nodes in top 10\% rank in degree )

\vspace{-0.12in}
\section{Conclusion and Future Work}
In this paper, we have proposed {\it SMART}, a novel framework for modeling and keeping track of spatial and temporal statistics of vehicle-to-infrastructure communication latency across a large geographical area. {\it SMART} can be directly applied to characterize the feasibility of supporting different latency-sensitive services  across a large geographical area during different time periods. Specifically, {\it SMART} first formulates the spatio-temporal performance and correlations of a vehicular network as a graphical structure and then adopt GCN and DQN to reconstruct the spatial and temporal latency performance in a slotted process. Simulation results show that the proposed method can improve both the modeling accuracy and reconstruction efficiency for large vehicular networks. 

Our work opens several potential directions that worth further investigating. In particular, it will be promising to extend {\it SMART} into a more general setting. In addition, it is also interesting to consider some other information to be included into the edge weights and input feature vector of GCN that can capture more complex correlations between vertices.

\vspace{-0.2in}
\section*{Acknowledgment}
This work was supported in part by the National Natural Science Foundation of China under Grants 62071193 and 61632019, the Key R \& D Program of Hubei Province of China under Grant 2020BAA002, and China Postdoctoral Science Foundation under Grant 2020M672357.

\vspace{-0.12in}
\bibliographystyle{IEEEtran}
\bibliography{bibljt}
\end{document}